\documentclass[11pt]{article}
\pdfoutput=1
\usepackage{jheppub}
\usepackage{amsmath}
\usepackage{bm}
\usepackage{slashed}
\usepackage{graphicx}

\newlength{\dummysp}
\newcommand{\beq}{\begin{eqnarray}}
\newcommand{\eeq}{\end{eqnarray}}

\newcommand{\gappeq}{\mathrel{\rlap {\raise.5ex\hbox{$>$}}
{\lower.5ex\hbox{$\sim$}}}}
\newcommand{\lappeq}{\mathrel{\rlap{\raise.5ex\hbox{$<$}}
{\lower.5ex\hbox{$\sim$}}}}

\newcommand{\ben}{\begin{enumerate}}
\newcommand{\een}{\end{enumerate}}

\newcommand{\bit}{\begin{itemize}}
\newcommand{\eit}{\end{itemize}}

\def\[{\left [}
\def\]{\right ]}
\def\({\left (}
\def\){\right )}
\def\R{{\mathbb R}}
\def\S{{\mathbb S}}
\def\Z{{\mathbb Z}}
\def\T{{\mathbb T}}

\title{Nonperturbative effects in the Standard Model  with gauged 1-form symmetry}
   
 \author[a]{Mohamed M. Anber,}\author[b]{Erich Poppitz} 
  
\affiliation[a]{Centre for Particle Theory, Department of Mathematical Sciences, Durham University, South Road, Durham DH1 3LE, UK}
\affiliation[b]{Department of Physics,   University of Toronto, 60 St George St., 
Toronto, ON M5S 1A7, Canada}
\emailAdd{mohamed.anber@durham.ac.uk}\emailAdd{poppitz@physics.utoronto.ca}    

\abstract{

{\flushleft{W}}e study the Standard Model with gauged $\mathbb Z_{N=2,3,6}^{(1)}$ subgroups of its  $\mathbb Z_6^{(1)}$ 1-form global symmetry, making the gauge group $SU(3) \times SU(2)\times U(1) \over \mathbb Z_N$. We show that, on a finite $\T^3$, there are self-dual instantons of fractional topological charge. They mediate baryon- and lepton-number violating processes. We compare their amplitudes to the ones due to the usual BPST-instantons. We find that the small hypercharge coupling suppresses the fractional-instanton contribution, unless the torus size is taken sub-Planckian, or extra matter is added above the weak scale. We also discuss these results in light of the cosmological bounds on the torus size. 
}

\begin{document}

\maketitle

\flushbottom

\section{Introduction, summary, and outlook.}

 It is well established that the Lie algebra of the Standard Model (SM) is $g_{SM}= su(3)\times su(2) \times u(1)$. The fact that there is freedom to choose different gauge groups $G_{SM}$  is less broadly appreciated \cite{Aharony:2013hda,Tong:2017oea}. If the gauge group is chosen to be $G_{SM}^1 \equiv SU(3)\times SU(2) \times U(1)$, the SM with its known matter content has a 1-form\footnote{The superscript $G^{(q)}$, with $q=1$, is used to distinguish 1-form symmetries (with group $G$), acting on line operators, from ordinary 0-form  symmetries ($q=0$) acting on local operators. The modern introduction of higher-form symmetries is found in \cite{Gaiotto:2014kfa}. The recent interest in the global aspects of gauge theories and their higher-form symmetries arose due to the need to resolve puzzles related to the mapping of Wilson and 't Hooft line operators under $S$-duality of ${\cal{N}}=4$ super-Yang-Mills \cite{Aharony:2013hda}. Further interest is due to the importance of higher-form symmetries for the recently discovered \cite{Gaiotto:2014kfa,Gaiotto:2017yup,Gaiotto:2017tne}
 generalized 't Hooft anomalies.
} $\mathbb Z_{6}^{(1)}$  global symmetry.  The global structure of the SM gauge group can be changed upon gauging its various discrete subgroups, leading to possible gauge groups $G_{SM}^N = { SU(3)\times SU(2) \times U(1) \over \mathbb \Z_N}$, with $N=1,2,3,6$. 
 The global structure of the SM  was discussed in \cite{Tong:2017oea}, using the language of ``genuine line operators'' of \cite{Aharony:2013hda}. Refs.~\cite{Garcia-Etxebarria:2018ajm,Wang:2020iqc,Davighi:2019rcd,Wan:2020ynf,Wan:2019soo,Wang:2020xyo}  studied gauge-anomaly related aspects of $G_{SM}^N$, in particular showing the absence of global anomalies in the SM with $G_{SM}^{N}$. 
 
 There are various ways to exhibit the $\mathbb Z_6^{(1)}$ 1-form symmetry of the SM. 
 In this paper, we take an ``old-fashioned'' approach, initiated by 't Hooft \cite{tHooft:1979rtg,tHooft:1981sps}   in the study of confinement in Yang-Mills theory (see also \cite{vanBaal:1982ag,Gonzalez-Arroyo:1997ugn,vanBaal:2000zc} and the recent work \cite{Unsal:2021cch,Unsal:2020yeh}). 
 Our main goal is to begin a discussion of the dynamical effects due to the different choices of $G_{SM}^{N}$. To this end, we consider the SM  in a finite volume, taking space to be a three-torus\footnote{We remind the    reader  that ``cosmic topology'' is a subject with a long history, reviewed in  \cite{Lachieze-Rey:1995qrb}.}  $\T^3$. As we review below, the choice of  SM gauge group $G_{SM}^{N}$ is reflected in the boundary conditions on the gauge and matter fields imposed on the spatial $\T^3$, in the boundary conditions in the time direction, and in the way the sum over the sectors with different boundary conditions is performed  in the path integral. 
 
Our main point, using insight due to 't Hooft \cite{tHooft:1981nnx}, is to show that for $G_{SM}^N$, with $N \ne 1$, on a finite $\T^3$  there exist self-dual instanton solutions of fractional topological charges. They mediate processes where the baryon number changes by a nonzero amount $\Delta B$, with integer $\Delta B = \Delta L$ (see Table \ref{different choices of n hyper} below). We calculate the exponential suppression factor for such processes for the various choices of SM gauge group $G_{SM}^N$ and compare it with that for similar processes mediated by the usual BPST \cite{Belavin:1975fg}  instantons computed by 't Hooft \cite{tHooft:1976snw,tHooft:1976rip}. 
 Owing to the fact that these fractional instantons are much less understood\footnote{See \cite{GarciaPerez:2000aiw,Gonzalez-Arroyo:2019wpu} for relevant recent work.} than the BPST instantons, we use the constant field strength self-dual solutions of \cite{tHooft:1981nnx,vanBaal:1984ar}. We find that  due to the smallness of the hypercharge coupling, the fractional-instanton mediated processes are exponentially suppressed compared to the usual BPST instantons. This suppression can be overcome if the torus size is taken extremely small (sub-Planckian, with only SM content)  or if extra matter  is added to make the hypercharge coupling increase faster above the weak scale.
We also discuss our findings in light of the cosmological bounds on the $\T^3$ size\footnote{See  \cite{Lachieze-Rey:1995qrb} and, e.g.~\cite{Levin:1998qq,Bond:1997ym,Aslanyan:2013lsa} for more recent work.} and speculate on the possible cosmological relevance of the global choice $G_{SM}^N$.
 
We leave various interesting aspects for future work. The way the solutions of  \cite{tHooft:1981nnx} are modified in the broken phase of the SM has not yet been studied. It would also be of interest to study the generalization and extension of the construction of non-constant field strength fractional instantons of   \cite{GarciaPerez:2000aiw,Gonzalez-Arroyo:2019wpu} to  the SM case. A better control of these issues might help clarify the possible role of the fractional instantons at larger torus sizes.
 
 \section{The $\Z_6^{(1)}$ and its gauging via twisted boundary conditions}
 
In this Section, we explain how the $\Z_6^{(1)}$ 1-form symmetry of the SM naturally  arises upon formulating the theory on a spatial 3-torus. This formulation is a convenient starting point to study the fractional topological charge solutions. We also discuss  the gauging of the symmetry and its relation to the boundary conditions imposed on the gauge fields in the path integral.  (An equivalent description of the gauging of the 1-form symmetry takes the form of a coupling to a 2-form topological quantum field theory  \cite{Kapustin:2014gua} and also requires that the spacetime manifold has noncontractible 2-cycles.)

To begin, we denote the $SU(3) \times SU(2) \times U(1)$ gauge couplings $g_3, g_2$, and $g_1$, respectively. For use below, note that the relation between the $U(1)$ coupling and the usual hypercharge coupling $g_Y$ is $g_1 = {g_Y\over 6}$, because we take all $U(1)$ charges to be integer, due to the assumed compactness of the $U(1)$ gauge group. Explicitly, the  charge assignments of the various SM matter fields are standard, except that the hypercharge is multiplied by $6$:
\begin{equation}
\label{charges}
\begin{array}{c|ccc|cc}\text{field}&SU(3)& SU(2)&U(1)&U(1)_B& U(1)_L \cr \hline
q_L&\square&\square&1&{1\over 3}& 0\cr
l_L &\bf{1}&\square&-3&0&1 \cr
\tilde{e}_R&\bf{1}&\bf{1}&6&0&-1\cr
\tilde{u}_R&\overline\square&\bf{1}&-4&-{1\over 3}&0\cr
\tilde{d}_R&\overline\square&\bf{1}&2&-{1\over 3}&0\cr
h&\bf{1}&\square&3&0&0\end{array}~.
\end{equation}
All fields above (but the Higgs $h$) are left-handed Weyl spinors and we have not included a right-handed neutrino as its possible existence will play no role in our discussion. For use below, in the two r.h.~columns we also give the standard baryon and lepton number assignments. In (\ref{charges}), we only show a single generation of fermions.

When formulated on  a   manifold of nontrivial topology, a gauge theory is defined on coordinate patches covering the manifold, and the gauge fields on different patches are related by transition functions, which are gauge group elements. The matter fields on different patches are also related by the  transition functions taken in the relevant representation of the gauge group. Here, following \cite{tHooft:1979rtg,tHooft:1981sps,vanBaal:1982ag,Gonzalez-Arroyo:1997ugn}, we consider the case of a torus, which  can be described by only one patch, but with nontrivial boundary conditions on the fields. 
If we were only interested in a Hamiltonian formulation of the theory, we would consider a spatial three-torus $\T^3$. In a path integral formulation, we study the finite temperature partition function of the theory on a four-torus $\T^4 \simeq \T^3 \times S^1$, where $\S^1$ is interpreted as a thermal circle; if we wish to  approach the zero-temperature limit, we take the size of the circle to infinity. 

\subsection{The $\T^4$ boundary conditions: transition functions and gauge transformations}
\label{sec:bc1}

We shall now discuss how the $Z_6^{(1)}$ symmetry arises on $\T^4$. The $\T^4$ is covered by a single closed coordinate patch $0 \le x^\mu \le a^\mu$, $\mu=1,2,3,4$. This single-patch construction can be seen as a limit of a multiple patch covering of the $\T^4$, as is   explicitly discussed in \cite{vanBaal:1982ag}. The gauge fields on the different sides of the $\T^4$, e.g. at $x^\mu=0$ and $x^\mu=a^\mu$ are related by transition functions, gauge group elements $\Omega_\mu$, which can be thought of as defined on the $x^\mu = a^\mu$ sides of the $\T^4$; gauge transformations are also group elements  defined on the entire coordinate patch.

To state the boundary conditions, begin by considering an arbitrary $\mu\nu$ 2-plane of the torus of sides $a^{\mu}$ and $a^\nu$. The coordinates are  identified as $x^\mu \equiv x^\mu + a^\mu$. Thus, the 2-plane is a square $0 \le x^\mu \le a^\mu$, $0 \le x^\nu \le a^\nu$ (for fixed values of the other coordinates). The fields on opposite sides of the square are related by transition functions, as we now describe.
 To set our notation, let the gauge fields of the SM be $A_{(j)} = A_{(j) \; \mu} dx^\mu$, with $j=1,2,3$ denoting the $U(1), SU(2)$, and $SU(3)$ gauge fields, respectively. 
 Here $A_{(j)\; \mu} = A_{(j)\; \mu}^a T_{(j)}^a$, where $T_{(j)}^a$ are the corresponding group generators in the fundamental representation, tr$T_{(j)}^a T_{(j)}^b = {1\over 2}\delta^{ab}$, $j=2,3$. For $U(1)$, $T_{(1)}^a$ are replaced by the identity.  
 The gauge field action for the $j$-th gauge group is
 \begin{equation}\label{gaugeaction}
 S_{(j)} = {1 \over 4 g_{(j)}^2} \int d^4 x \;F_{(j)\; \mu\nu}^a F_{(j)}^{\mu\nu \; a}~,
\end{equation}
where for $j=1$ there is no sum over $a$.
 The covariant derivative acting on, e.g., $q_L$ with charges from (\ref{charges}), is then $D_\mu q_L =  
(\partial_\mu + i A_{(1)\; \mu} + i A_{(2)\; \mu}^a T_{(2)}^a + i A_{(3) \; \mu}^a T_{(3)}^a) q_L$. Gauge transformations of the $j$-th group act on the gauge fields as $A_{(j)} \rightarrow g_{(j)} (A_{(j)}- i d) g_{(j)}^{-1}$. On the matter fields, these are taken in the appropriate representation, e.g. $q_L \rightarrow g_{(j)} q_L$, where $g_{(j)} $ is a group element in the fundamental representation defined in the $0 \le x^\mu \le a^\mu$ coordinate patch ($\mu=1,...4$). For $U(1)$, the gauge group element is simply $g_{(1)} = e^{i \omega_{(1)}}$ when acting on a field of the minimal quantized unit charge.

The boundary conditions on the gauge and matter fields in the $\mu$-th direction of the square are then given by
\begin{eqnarray}\label{bc}
A_{(j)}(x^\mu = a^\mu) &=& \Omega_{(j)\; \mu} \left(A_{(j)}(x^\mu = 0)- i d\right) \Omega_{(j) \; \mu}^{-1}~, ~ j=1,2,3,~\\
\psi^{\cal{R}}(x^\mu = a^\mu)&=& \Omega_{(1) \;\mu}^{\cal{R}}\Omega_{(2) \;\mu}^{\cal{R}}
\Omega_{(3) \;\mu}^{\cal{R}}\; \psi^{\cal{R}}(x^\mu = 0)~, \text{where} ~ \psi^{\cal{R}} = (q_L, l_L, \tilde{e}_R, \tilde{u}_R, \tilde{d}_R, h)~.\nonumber
\end{eqnarray}
Above, we did not explicitly indicate the fact that the fields depend on all but the $x^\mu$ coordinates, i.e. their values are taken on the  3-planes $x^\mu=a^\mu$ or $x^\mu=0$ in $\T^4$.
The boundary conditions for the  $\nu$-th direction are identical, with $\mu \rightarrow \nu$, and are given in terms of their own transition functions $\Omega_{(j)\; \nu}$. 

{\flushleft{L}}et us now make some comments regarding the boundary conditions (\ref{bc}):
\begin{enumerate}
\item
The notation in (\ref{bc}) may be somewhat condensed, so let us elaborate. The functions $\Omega_{(j) \; \mu}$ are the transition functions in the $\mu$-th direction of the torus for the $j$-th gauge group. They are group elements of the  $j$-th gauge group in the fundamental representation. Note  that the transition functions  $\Omega_{(j) \; \mu}$ in the $\mu$-th direction are  independent of $x^\mu$ and can be thought to be defined at  the $x^\mu = a^\mu$ side of the torus. The matter fields at $x^\mu = a^\mu$ and $x^\mu = 0$ are related as shown in the second line in (\ref{bc}), where $\Omega_{(j) \; \mu}^{\cal{R}}$ are the transition functions taken in the appropriate representation from (\ref{charges}): $\Omega_{(j) \; \mu}^{\cal{R}}$ are the same group elements  $\Omega_{(j) \; \mu}$, but taken in the representation ${\cal{R}}$ of the fermion $\psi^{\cal{R}}$. Explicitly, from eqn.~(\ref{charges}), we see that for the SM and for $j=2,3$ only the fundamental and antifundamental representations appear, hence $\Omega_{(j) \; \mu}^{\cal{R}} = \Omega_{(j) \; \mu}$ if the field $\psi^{\cal{R}}$ is in the fundamental representation or $\Omega_{(j) \; \mu}^{\cal{R}} =\Omega_{(j) \; \mu}^*$ if it is in the antifundamental representation. For a $\psi^{\cal{R}}$ field of charge $q_1$ under $U(1)$, we have $\Omega_{(1) \; \mu}^{{\cal{R}} = q_1} = (\Omega_{(1) \; \mu})^{q_1}$, where $\Omega_{(1)\; \mu} = e^{i \omega_{(1)\; \mu}}$ is a compact-$U(1)$ group element. It is clear that with this definition, the matter field and its covariant derivative obey the boundary conditions (\ref{bc}).
\item
Consistency of the boundary conditions (\ref{bc}) with gauge transformations of each gauge group   requires that the transition functions $\Omega_{(j)\; \mu}$ themselves transform under gauge transformations: $\Omega_{(j)\; \mu} \rightarrow g_{(j)}(x^\mu = a^\mu) \;\Omega_{(j)\; \mu} \; g_{(j)}^{-1}(x^\mu = 0)$, and similar for the matter fields, where $\Omega$ and $g$ are to be taken in the appropriate representation.  For example, when studying classical solutions, one can make use of this freedom to choose a convenient form of the transition functions. Using this gauge freedom to, e.g., set some transition functions to unity, amounts to partial gauge fixing (see below). 
\item
Most important for our further discussion is the fact that the transition functions obey  consistency conditions in every $\mu\nu$-plane of the torus. These are a consequence of the  well-known ``cocycle conditions'' on general manifolds and we shall also call them by that name. In our single-chart setup 
 the cocycle conditions follow by requiring that the fields are single valued on every $\mu\nu$-square.  To make the point, consider the matter field $\psi^{\cal{R}}$. The value of $\psi^{\cal{R}}$ at the ($x^\mu =a^\mu, x^\nu = a^\nu$) corner of the square can be obtained from its value at the opposite ($x^\mu = x^\nu=0$) corner by using the boundary conditions (\ref{bc}) in two different ways: first go along the $x^\mu=0$ side in the $\nu$-th direction and then along the $x^\nu=a^\nu$ side in the $\mu$-th direction,  or v.v. Obtaining the same value of the $\psi^{\cal{R}}$ field at the corner then requires that
  \begin{eqnarray}
 \label{cocycle1}
&&  \Omega_{(1) \;\mu}^{\cal{R}} \Omega_{(2) \;\mu}^{\cal{R}} 
\Omega_{(3) \;\mu}^{\cal{R}}(x^\nu=a^\nu)  \times  \Omega_{(1) \;\nu}^{\cal{R}} \Omega_{(2) \;\nu}^{\cal{R}}
\Omega_{(3) \;\nu}^{\cal{R}} (x^\mu=0) \\
=&&    \Omega_{(1) \;\nu}^{\cal{R}} \Omega_{(2) \;\nu}^{\cal{R}}
\Omega_{(3) \;\nu}^{\cal{R}} (x^\mu=a^\mu) \times   \Omega_{(1) \;\mu}^{\cal{R}} \Omega_{(2) \;\mu}^{\cal{R}} 
\Omega_{(3) \;\mu}^{\cal{R}}(x^\nu=0) \nonumber ~.
\end{eqnarray}
This should hold for all $\mu, \nu$ planes and all $\psi^{\cal{R}}$ from (\ref{charges}). 

\item
There are consistency conditions similar to (\ref{cocycle1}) for the $\Omega_{(j) \; \mu}$ appearing in the gauge-field boundary conditions. It turns out that there is some freedom in the gauge field cocycle conditions that we describe below. In fact, the different $G_{SM}^N$ theories are distinguished by the conditions obeyed by the transition functions for the gauge fields as we now discuss.
\end{enumerate}

\subsection{$G_{SM}^1$ and the $\Z_6^{(1)}$ global symmetry}

Here, we discuss the cocycle conditions on the transitions functions for the gauge field and introduce the $\Z_6^{(1)}$ symmetry of the SM.

Using the same logic as for the field $\psi^{\cal{R}}$ to obtain (\ref{cocycle1}), we see that to obtain a single-valued gauge field, it is sufficient\footnote{But not necessary: running ahead, we immediately notice that because  the gauge field boundary condition in (\ref{bc}) contains both $\Omega$ and $\Omega^{-1}$, one can allow a phase factor equal to a center element of each gauge group to appear on the r.h.s. of (\ref{cocycle2}). } to take 
the gauge field transition functions to obey a condition identical to (\ref{cocycle1}) for each $j=1,2,3$. We call this cocycle condition the $G_{SM}^1$ condition:
\begin{eqnarray}
\label{cocycle2}
G_{SM}^1:~~  \Omega_{(j) \;\mu} (x^\nu=a^\nu)   \;  \Omega_{(j) \;\nu} (x^\mu=0)=  \Omega_{(j) \;\nu} (x^\mu=a^\mu) \;\Omega_{(j) \;\mu} (x^\nu=0) ~, ~j=1,2,3.
\end{eqnarray}
The fields summed over in the path integral for $G_{SM}^1$ satisfy the boundary conditions (\ref{bc}) with transition functions obeying the cocycle conditions (\ref{cocycle1}, \ref{cocycle2}). Of  particular interest to us is the fact that the topological charge for $G_{SM}^1$ is integer \cite{tHooft:1979rtg,tHooft:1981sps,vanBaal:1982ag}, to be discussed below.\footnote{It can be shown that, using the gauge freedom discussed in Section \ref{sec:bc1}, one can take the $j=2,3$ transition functions obeying (\ref{cocycle2}) to be trivial in  the spatial $\T^3$ directions, $\Omega_{(j)\; \mu} = 1$ for $\mu=1,2,3$,  and only have a nontrivial $x^4$-direction transition function $\Omega_{(j) \; 4} \ne 1$. In this gauge, the transition function in the ``time'' direction carries all the information about the integer topological charge.}

  Now we can explicitly exhibit the 1-form $\Z_6^{(1)}$ center symmetry of the SM. This is a $\Z_6$ global  symmetry, parameterized by two integers, $k_\mu^2\in \Z (\text{mod}\, 2)$ and $k_\mu^3\in \Z (\text{mod}\, 3)$, for every spacetime direction $\mu$. In this formulation, the 1-form center symmetry does not act on any fields, but acts only on the transition functions as follows
\begin{eqnarray}
\label{z61}
Z_6^{(1)}:~\Omega_{(1) \; \mu} &\rightarrow& e^{i 2 \pi (- {k_\mu^2 \over 2} - {k_\mu^3 \over 3})} \; \Omega_{(1) \; \mu}~,~ k_\mu^2 \in \{ 1,2\},\; k_\mu^3 \in \{ 1,2,3\}, \; \mu = 1,2,3,4,\nonumber \\
\Omega_{(2) \; \mu} &\rightarrow& e^{i 2 \pi {k_\mu^2 \over 2} } \; \Omega_{(2) \; \mu}~, \\
\Omega_{(3) \; \mu} &\rightarrow& e^{i 2 \pi  {k_\mu^3 \over 3}} \; \Omega_{(3) \; \mu}~. \nonumber 
\end{eqnarray}
Clearly the action (\ref{z61}) is consistent with  $\Omega_{(j) \; \mu}$ being (special unitary, for $j=2,3$) group elements. 

That the above $\Z_6^{(1)}$ is a symmetry follows from the fact that both the boundary conditions (\ref{bc}) and the gauge and matter fields' cocycle conditions (\ref{cocycle1}), (\ref{cocycle2}) are invariant under (\ref{z61}).
  The gauge field boundary condition, the first line in (\ref{bc}), is clearly invariant since both $\Omega_{(j)\; \mu}$ and $\Omega_{(j)\; \mu}^{-1}$ appear on the r.h.s. The cocycle  conditions for the gauge field (\ref{cocycle2}) are also  obviously invariant, as $\Omega_{(j) \; \mu}$ appears on both sides of the equations.  The matter field boundary conditions from the second line in (\ref{bc}) are also invariant, for all $\psi^{\cal{R}}$ from (\ref{charges}), as is easily  checked explicitly and follows from the charge assignments. For example, for $q_L$ this follows immediately form the fact that the phase factors  on the r.h.s. of  (\ref{z61}) simply cancel out. That the cocycle conditions for the matter fields (\ref{cocycle1}) are also invariant follows from the invariance of the boundary conditions (\ref{bc}) for $\psi^{\cal{R}}$, as the products $\Omega_{(1) \;\mu}^{\cal{R}} \Omega_{(2) \;\mu}^{\cal{R}} 
\Omega_{(3) \;\mu}^{\cal{R}}$ are invariant under (\ref{z61}) for all $\cal{R}$.

 We have not yet explained why the symmetry (\ref{z61}) is a 1-form symmetry, in the modern terminology of \cite{Gaiotto:2014kfa}: this is because the only gauge invariant operators that (\ref{z61}) acts on are Wilson lines winding around the noncontractible loops of the torus. It is clear that all local gauge invariant operators are blind to a symmetry like (\ref{z61}) that only acts on transition functions (since a local operator can always be written using fields within a given coordinate patch). On the other hand, a line operator, such as the fundamental representation Wilson loop of the $j$-th gauge group, winding once around the $x^\mu$ direction, has the form
\begin{equation}\label{wilson}
W_{(j) \; \mu} = \text{tr} ( {\cal{P}} \;e^{ \; i \int\limits_{x^\mu=0}^{x^\mu=a^\mu} A_{(j)} } \;\Omega_{(j) \; \mu} )~,
\end{equation}
where $\cal{P}$ denotes path ordering and the transition function is inserted in the trace in order  to ensure the invariance of $W_{(j) \; \mu}$  under gauge transformations $g_{(j)}$ on the $\T^4$ (recall the gauge transformation law of $\Omega_{(j) \; \mu}$ mentioned above). It is now clear that under (\ref{z61}), we have that 
\begin{eqnarray}\label{wilson2}
W_{(3) \; \mu} &\rightarrow& e^{i 2 \pi  {k_\mu^3 \over 3}} \; W_{(3) \; \mu}~,  \\
W_{(2) \; \mu} &\rightarrow& e^{i 2 \pi  {k_\mu^2 \over 2}} \; W_{(2) \; \mu}~,  \\
W_{(1) \; \mu} &\rightarrow& e^{i 2 \pi (- {k_\mu^2 \over 2}- {k_\mu^3 \over 3})} \; W_{(1) \; \mu}~,
\end{eqnarray}
showing that $\Z_6^{(1)}$ acts on the above fundamental Wilson loops (thus, they are genuine line operators in the SM, as discussed in great detail in \cite{Tong:2017oea}).\footnote{We also note that the introduction of the 1-form symmetry via its action on transition functions is closest to the way the global $\Z_N$ center-symmetry is introduced in $SU(N)$ lattice gauge theory, as described in \cite{Greensite:2011zz}.}

 \subsection{The gauging of $\Z_N^{(1)} \in \Z_6^{(1)}$ and $G_{SM}^{N=2,3,6}$ }
Having explained the way the global $\Z_6^{(1)}$ 1-form symmetry appears in the SM in this language, we can now introduce the different $G_{SM}^N$ theories obtained upon the gauging of the  $\Z_{N=2,3,6}^{(1)}$ subgroups of $\Z_6^{(1)}$. We describe the gauging using the $\T^4$ formulation of the gauge theory, as the fractional instantons whose role we want to study are only explicitly known for $\T^4$.
 
 When  the gauge theory is formulated  on $\T^4$ in the Euclidean  path integral formalism, the integral is performed over all  gauge fields and transition functions obeying the cocycle conditions (\ref{cocycle1}) and (\ref{cocycle2}), modulo gauge transformations. The cocycle conditions on the gauge field  transition functions in (\ref{cocycle2}) are the ones appropriate to the $G_{SM}^1 = SU(3) \times SU(2) \times U(1)$ gauge group. As already mentioned, with these cocycle conditions, 
the topological charge is an integer (see (\ref{charge23}) and (\ref{charge1}) below).

Now we come to the gauging of the 1-form symmetry. Adapting the discussion of  \cite{Kapustin:2014gua} to our case, 
the gauging of subgroups of $\Z_6^{(1)}$ involves two steps: a modification of the gauge fields' cocycle conditions (\ref{cocycle2}) by appropriate center elements and a summation in the path integral over gauge fields obeying all such modified cocycle conditions (possibly including $\Z_N$ phase factors called ``discrete theta angles,'' introduced in \cite{Aharony:2013hda} and described for $G_{SM}^N$ in \cite{Tong:2017oea}). 

We begin with modified gauge-field cocycle conditions (\ref{cocycle2}) for the SM case:
 \begin{eqnarray}
 \label{cocycle3}
 \Omega_{(3) \;\mu} (x^\nu=a^\nu)   \;  \Omega_{(3) \;\nu} (x^\mu=0)&=& e^{i 2 \pi {n_{\mu\nu}^{(3)} \over 3}} \; \Omega_{(3) \;\nu} (x^\mu=a^\mu) \;\Omega_{(3) \;\mu} (x^\nu=0)~,\nonumber \\
  \Omega_{(2) \;\mu} (x^\nu=a^\nu)   \;  \Omega_{(2) \;\nu} (x^\mu=0)&=& e^{i 2 \pi {n_{\mu\nu}^{(2)} \over 2}} \;\Omega_{(2) \;\nu} (x^\mu=a^\mu) \;\Omega_{(2) \;\mu} (x^\nu=0)~, \\
    \Omega_{(1) \;\mu} (x^\nu=a^\nu)   \;  \Omega_{(1) \;\nu} (x^\mu=0)&=& e^{i 2 \pi(- {n_{\mu\nu}^{(3)} \over 3}- {n_{\mu\nu}^{(2)} \over 2})}\; \Omega_{(1) \;\nu} (x^\mu=a^\mu) \;\Omega_{(1) \;\mu} (x^\nu=0)~. \nonumber
 \end{eqnarray}
 The $x$-independent integers appearing in the cocycle conditions for $SU(2)$ transition functions, $n_{\mu\nu}^{(2)}= - n_{\nu\mu}^{(2)}$, are defined (mod $2$) for every 2-plane, while the ones for $SU(3)$, $n_{\mu\nu}^{(3)}= - n_{\nu\mu}^{(3)}$, are similarly defined (mod $3$). The $n_{\mu\nu}^{(2),(3)}$ are known as 't Hooft twists. In the formalism introduced in \cite{Kapustin:2014gua} (which we shall not elaborate upon) the integers $n_{\mu\nu}^{(2),(3)}$ correspond to the integrals,   taken over the various 2-planes of the torus, of the appropriate 2-form discrete gauge fields gauging the 1-form $\Z_N^{(1)}$ symmetry of the $SU(N)$ theory (here $N=2,3$). The summation over field configurations obeying (\ref{cocycle3}) with $n_{\mu\nu}^{(2)}=\{1,2\}$ and $n_{\mu\nu}^{(3)}=\{1,2,3\}$ amounts to performing the path integral over this topological field.\footnote{In a Hilbert space interpretation (whose details  we shall also not need) this is a summation over discrete magnetic flux sectors and a projection on appropriate discrete electric flux sectors. These sectors were introduced in \cite{tHooft:1979rtg,tHooft:1981sps}. See also \cite{Aharony:2013hda,Cox:2021vsa} for somewhat related discussions.}
  
  The twists  of the gauge field cocycle conditions in (\ref{cocycle3}) are consistent with the  cocycle conditions (\ref{cocycle1}) of the  SM matter fields $\psi^{\cal{R}}= (q_L, l_L, \tilde{e}_R, \tilde{u}_R, \tilde{d}_R, h)$,  as the phase factors cancel out in the products  $\Omega_{(1) \;\mu}^{\cal{R}} \Omega_{(2) \;\mu}^{\cal{R}} 
\Omega_{(3) \;\mu}^{\cal{R}}$ for all representations $\cal{R}$ in (\ref{charges}). Once again, this is most obvious for $q_L$, whose cocycle condition (\ref{cocycle1}) is satisfied by the product of the three lines in (\ref{cocycle3}), and is easily verified for the other SM matter fields.

The upshot is that the different $G_{SM}^N$ gauge theories are distinguished by the values of the twists of the cocycle conditions (\ref{cocycle3}) obeyed by the transition functions of the gauge and matter fields  summed over in the path integral:\footnote{To avoid confusion, we stress that one can, even in the $G_{SM}^1$ theory, introduce boundary conditions on the $\T^4$ that correspond to nonzero twists $n_{\mu\nu}^{(j)}$, but without summation over all allowed twists in the path integral. This corresponds to probing the $G_{SM}^1$ theory by the insertion of nondynamical backgrounds for the $\Z_6^{(1)}$ global symmetry, needed in studies of generalized 't Hooft anomalies involving center symmetry, as shown using this language in \cite{Cox:2021vsa}. An unambiguous identification of the different SM gauge groups with the presence or absence of fractional instanton contributions requires forbidding such nondynamical insertions. This is natural in the context of embedding the SM in quantum gravity theories, where global symmetries cannot appear; for example, in the case of $G_{SM}^{(1)}$, the $\Z_6^{(1)}$ symmetry is broken by the presence of    heavy electric states whose transition functions do not respect (\ref{z61}), disallowing nonzero twists in $G_{SM}^{(1)}$.  Likewise, global $\Z_3^{(1)}$ twists can be forbidden in $G_{SM}^{(2)}$ and $\Z_2^{(1)}$ ones in $G_{SM}^{(3)}$. It is also worth mentioning that there is a $\mathbb Z_3$ magnetic symmetry in $G_{SM}^{(3)}$ and $G_{SM}^{(6)}$, which is broken by heavy magnetic charges in theories of quantum gravity. See \cite{Tong:2017oea} and references therein for a discussion of these issues in the present context.}
 \begin{eqnarray}\label{gaugingrule}
 G_{SM}^1 = SU(3)\times SU(2) \times U(1): && n_{\mu\nu}^{(3)} \equiv 0,  n_{\mu\nu}^{(2)} \equiv 0, \; \forall \mu, \nu, \nonumber\\
 G_{SM}^2 = {SU(3)\times SU(2) \times U(1) \over \Z_2}: && n_{\mu\nu}^{(2)}= \{1,2\}, \text{with} ~n_{\mu\nu}^{(3)}  \equiv 0 \; \forall \mu, \nu, \nonumber\\
 G_{SM}^3 = {SU(3)\times SU(2) \times U(1) \over \Z_3}: && n_{\mu\nu}^{(3)} = \{1,2, 3\}, \text{with} ~n_{\mu\nu}^{(2)} \equiv 0 \; \forall \mu, \nu,\nonumber\\
 G_{SM}^6 = {SU(3)\times SU(2) \times U(1) \over \Z_6}: && n_{\mu\nu}^{(2)}= \{1,2\}, n_{\mu\nu}^{(3)} = \{1,2, 3\} .
\end{eqnarray}

From our point of view, the most important effect of the gauging of $\Z_6^{(1)}$ subgroups in the SM is the fact that the topological charge of gauge fields obeying the modified cocycle conditions (\ref{cocycle2}) can be fractional. It was first argued by 't Hooft in 
 \cite{tHooft:1979rtg,tHooft:1981sps}, and rigorously shown in  \cite{vanBaal:1982ag}, that the fractional part of the topological charge is determined by the twists $n_{\mu\nu}^{(2),(3)}$. In particular, the topological charge for the $j$-th gauge group  is (recalling $F_{(j)} ={1\over 2} F_{(j) \; \mu\nu} dx^\mu \wedge dx^\nu$ and noting that there is no trace for $j=1$):
 \begin{eqnarray}\label{topcharge}
 Q_{(j)} =  {1 \over 8 \pi^2}\int  \; \text{tr} F_{(j)} \wedge F_{(j)} =\left\{ \begin{array}{c} ~~ {1 \over 32 \pi^2}\int  d^4 x \;F_{(j)}^{\mu\nu \; a} \; \tilde F_{(j) \; \mu\nu}^{a}, ~j=2,3,~\cr \cr{1 \over 16 \pi^2}  \int d^4 x \;F_{(1)}^{\mu\nu}\; \tilde F_{(1) \; \mu\nu},~j=1,\end{array}\right.~
 \end{eqnarray}
 where $\tilde F_{(j)\; \mu\nu}\equiv {1 \over 2} \epsilon_{\mu\nu\lambda\sigma} F_{(j)}^{\lambda\sigma}$.
 For $j=2$ and $j=3$, it was shown in \cite{tHooft:1979rtg,tHooft:1981sps,vanBaal:1982ag}
that $SU(j)$ field configurations obeying the cocycle conditions (\ref{cocycle3}) have fractional topological charge
\begin{equation}\label{charge23}
Q_{(j)} = - {\text{Pf}(n^{(j)}) \over j} + k~,~ j=2,3, ~ \text{and} ~k\in \Z~.
 \end{equation}
 Here, $\text{Pf}(n) \equiv{1\over 8} \epsilon^{\mu\nu\lambda\sigma} n_{\mu\nu} n_{\lambda\sigma}$ is an integer for an antisymmetric matrix with integer entries. The simplest example is to take $n_{12} = - n_{21}=n_{34} = - n_{34} = 1$ as the only nonzero entries of $n_{\mu\nu}$, giving $\text{Pf}(n)= n_{12} n_{34} = 1$. 
 
We used eqn.~(\ref{topcharge}) to define the  quantity $Q_{(1)}$ characterizing the  $U(1)$ gauge field configuration, noting that there is no $U(1)$ characteristic class associated with the $\T^4$. Instead, the $U(1)$ background is characterized by the fluxes through noncontractible 2-planes (``first Chern class''). With nontwisted cocycle conditions for the $U(1)$, i.e. with $U(1)$ transition functions obeying (\ref{cocycle2}),  field configurations are characterized by $2\pi \times$integer fluxes through the various 2-planes, e.g. $\oint dx^1 dx^2 F_{(1)\; 12} = 2 \pi \Z$. It is easy to see that our $Q_{(1)}$ of (\ref{charge23}) then takes  integer values  (see eqn.~(\ref{charge1}) below).
 For $U(1)$ field configurations obeying the cocycle conditions from the last line in (\ref{cocycle3}), which are modified by the $\Z_6$ twists $\bar n_{\mu\nu}$ (with ${\bar n_{\mu\nu}\over 6} \equiv {-2 n_{\mu\nu}^{(3)} - 3n_{\mu\nu}^{(2)} \over 6}$), these fluxes through 2-planes are now valued in $2\pi \Z\over 6$ and $Q_{(1)}$ is further reduced, as also seen in (\ref{charge1}) below.

 The fact that the topological charge is fractional leads one to expect that the physical effects of such topological fluctuations may be less suppressed than the ones due to the well-understood integer topological charge BPST instantons. The gauge field Euclidean action (\ref{gaugeaction}) can be rewritten in terms of the topological charge $Q_{(j)}$ (for $j=2,3$) as
 \begin{equation}
 \label{the full action}
 S_{(j)} = \pm {8 \pi^2 \over g_{(j)}^2} Q_{(j)} + {1 \over 8 g_{(j)}^2} \int d^4 x (F_{(j)\; \mu\nu} \mp \tilde F_{(j)\; \mu\nu})^2
  \end{equation}
 where the signs are correlated, showing that the action for field configurations with $Q_{(j)} >0$  are minimized by  
self-dual fields $ F_{(j)\; \mu\nu} = \tilde F_{(j)\; \mu\nu}$ (and anti self-dual ones for negative topological charge). Thus, the minimal action in each topological sector $Q_{(j)}$ (for $j=2,3$) is given by the familiar Bogomolny' bound 
\begin{equation}\label{bound}
S_{min} = {8 \pi^2 \over g_{(j)}^2} |Q_{(j)}| = {8 \pi^2 \over g_{(j)}^2}|{{\text{Pf}}(n^{(j)}) \over j}-k|~.
\end{equation} For example, with $|Q_{(j)}| = {1 \over j}$, the action is smaller than that of a BPST instanton. In particular, solutions whose action scales as   $8 \pi^2/(g^2 N)$, as  above, with $j \rightarrow N$, give contributions to the path integral which survive in the large-$N$ limit---which was the initial motivation of their study in \cite{tHooft:1981nnx}.

 Before we continue, we recall the anomaly equation for the $U(1)_B$ baryon number of (\ref{charges}), 
 \begin{equation}
 \label{anomalyb}
 \partial_\mu j^\mu_B =  {1 \over 32 \pi^2}\ F_{(2)}^{\mu\nu \; a} \: \tilde F_{(2)\; \mu\nu}^{a}- 18  \;{1 \over 16 \pi^2}\ F_{(1)}^{\mu\nu }\; \tilde F_{(1) \; \mu\nu },  \end{equation}
 which implies that the change of baryon number is $\Delta B = Q_{(2)} - 18 Q_{(1)}$, for a single generation of SM fields.
In the next Section, we describe the $G_{SM}^N$ fractional instantons and study their effect.

  \section{The fractional instantons of $G_{SM}^N$, their actions, and the $\Delta B \ne 0$ amplitudes}
 
 \subsection{General remarks and concerns}
 
 Before we exhibit the fractional instantons for the $G_{SM}^N$ theories that we shall use as our illustrative example, we note that the current knowledge of such solutions is somewhat limited.  The fact that the  topological charge is fractional has been known since \cite{tHooft:1979rtg,tHooft:1981sps,vanBaal:1982ag}, but  there are only a few explicit  solutions. 

Below, as a first step to investigating the possible role of fractional instantons, we make two simplifications, both of which should be reconsidered in future studies. First, we set the vev of the Higgs field to zero assuming a symmetric phase.   Second,  we  use the constant field strength solutions on $\T^4$ of  \cite{tHooft:1981nnx}, and,  in particular,  their abelian version.  These analytic solutions saturate the Bogomolny' bound (\ref{bound}) and are stable only for  a particular ratio of sides of the $\T^4$   \cite{vanBaal:1984ar}. In particular, using these solutions, we are forced to take the $\T^4$ sides   to obey \begin{equation}\label{torussize}
{a^3 a^4 \over a^1 a^2} = 1~.
\end{equation} 
This constraint ensures that both the $j=2$ and $j=3$ fractional instantons are self-dual and obey the Bogomolny' bound (\ref{bound}). The $SU(3)$ action of these solutions, however, is only $2/3$ the BPST instanton action (rather than the minimal\footnote{\label{footnoteabelian}We note that  't Hooft \cite{tHooft:1981nnx} constructed $SU(N_c=3)$ nonabelian constant field strength solutions of the minimal possible action (\ref{bound}), equal to $1/3$ times the BPST action. However, the relation   replacing (\ref{torussize}) and ensuring self-duality then has the ratio $a^3 a^4/(a^1 a^2)$   replaced with a quantity $f(N_c,k,l)$ (with $k+l=N_c$) that depends on $N_c$. Hence, self duality is not obeyed for both $N_c=2$ and $N_c=3$, prompting us to use the abelian version. The generalization of these solutions to product-group  structures such as the SM has not been studied.} $1/3$), while the $SU(2)$ action is the minimum possible $1/2$ of the BPST action.
 
 Clearly, the the constraint (\ref{torussize}) is a source of concern. First,  it limits the ability to independently vary the shape and size of the spatial $\T^3$ and  of the time direction extent $a_4$. In particular, it does not incorporate zero-temperature fractional instantons on $\R \times \T^3$  \cite{vanBaal:2000zc}. 
Further, since the solutions have   field strength uniform over the torus, one expects them to be relevant for possible physical applications only if the size of the torus is smaller than the causally connected part of the universe.  As we discuss below, this fact, along with the cosmological bounds on the torus size, would then imply that the SM fractional instantons are mostly of academic interest. 

 While this may ultimately turn out to be the case, let us make several remarks pointing towards a possible way out.  There exist numerical studies as well as an analytic approach towards finding fractional instantons on tori not restricted by (\ref{torussize}) or its generalization. More precisely, fractional instanton solutions can be constructed by relaxing the 
ratio ${a^3 a^4\over a^1 a^2}=f(N,k,l)$, with $k+l=N$ ('t Hooft's    generalization of  (\ref{torussize})) towards arbitrary values of $a^\mu$, using a perturbative approach, developed in \cite{GarciaPerez:2000aiw,Gonzalez-Arroyo:2019wpu}. It is shown there that the fractional instanton solutions  thus obtained have a nontrivial profile of the field strength over the torus. The instanton size is, however,   still set by the torus dimensions (the equations studied are the scale-free Yang-Mills equations on a torus with twisted boundary conditions, as in the top  line of (\ref{cocycle3})).  It appears, then, that fractional instantons whose size does not scale with the size of space (time) would require the introduction of an extra scale. This could be related to the strong scale of QCD or to the electroweak scale,\footnote{Notice that  it is precisely the appearance of an extra scale, due to the holonomy vev, that reveals the fractional instantons in the much better understood case of $\R^3 \times \S^1$ \cite{Lee:1997vp,Kraan:1998pm}.
} the Higgs vev that we neglect throughout. It is thus of interest to study especially (in the present context) the  effect of re-introducing the Higgs vev.  

 With our  concerns spelled out, we now describe the fractional instantons on $\T^4$. 
 
 \subsection{The fractional instantons in $G_{SM}^N$}
 To describe the fractional instantons explicitly, we introduce $H^{(3)}$ and $H^{(2)}$,  diagonal matrices in the Cartan subalgebra  of  $SU(3)$ and $SU(2)$:
 \begin{eqnarray}\label{hmatrices}
 H^{(3)} = \text{diag}({1\over 3}, {1\over 3}, -1 + {1 \over 3}),~~H^{(2)} = \text{diag}({1\over 2}, - {1 \over 2}), ~ \text{tr} H^{(3)} H^{(3)} = {2 \over 3}, ~ \text{tr} H^{(2)} H^{(2)} = {1 \over 2}. 
 \end{eqnarray}
The fractional instantons are the following field configurations, where $n^{(3),(2)}_{\mu\nu}$ are the  twists appearing in (\ref{cocycle3}), as we show below:
 \begin{eqnarray}\label{fractionalinstantons}
 A_{(j)} &=& \sum\limits_{\mu,\nu=1}^4 {\pi x^\nu dx^\mu n_{\mu\nu}^{(j)} \over a^\mu a^\nu} H^{(j)}, ~ \text{for}~ j= 2,3,\nonumber\\
 A_{(1)} &=& - \sum\limits_{\mu,\nu=1}^4 {\pi x^\nu dx^\mu \over a^\mu a^\nu} \left( {3 n_{\mu\nu}^{(2)} + 2 n_{\mu\nu}^{(3)} \over 6} + n^{(1)}_{\mu\nu}\right)~.
 \end{eqnarray}
 In the last line, we introduced an integer-valued $n^{(1)}_{\mu\nu}=-n^{(1)}_{\nu\mu}$ twist reflecting the freedom to add $2\pi \Z$ fluxes of the $U(1)$ field.
Notice that the gauge fields (\ref{fractionalinstantons}) are not periodic, but instead obey
\begin{eqnarray}
 A_{(j)}(x^\lambda=a^\lambda) &-& A_{(j)}(x^\lambda=0) = ~~\sum\limits_{\mu}^4 {\pi  dx^\mu n_{\mu\lambda}^{(j)} \over a^\mu} H^{(j)} \equiv    - i \Omega_{(j) \; \lambda} d \Omega_{(j) \; \lambda}^{-1}\nonumber, ~\text{for}~ j= 2,3,\\
 \nonumber
 A_{(1)}(x^\lambda=a^\lambda) &-&A_{(2)}(x^\lambda=0) = - \sum\limits_{\nu=1}^4 {\pi   dx^\mu \over a^\mu } \left( {3 n_{\mu\lambda}^{(2)} + 2 n_{\mu\lambda}^{(3)} \over 6}+ n^{(1)}_{\mu\nu}\right)\equiv  - i \Omega_{(1) \; \lambda} d \Omega_{(1) \; \lambda}^{-1} ~,\\
 \end{eqnarray}
 from which we can read off the transition functions
 \begin{eqnarray}
 \label{omegabackgrd}
 \Omega_{(j) \; \lambda} =  e^{- i \sum\limits_{\mu}{\pi x^\mu n^{(j)}_{\mu\lambda} \over a^\mu} H^{(j)}},~ \text{for}~j=2,3, ~\text{while} ~
 \Omega_{(1) \; \lambda}  =  
e^{ i \sum\limits_{\mu}{\pi x^\mu \over a^\mu} \left( {3 n_{\mu\lambda}^{(2)} + 2 n_{\mu\lambda}^{(3)} \over 6}+ n^{(1)}_{\mu\nu}\right) }. 
 \end{eqnarray}
We can now ask what cocycle conditions  the above transition functions obey. To this end,  we   compute the l.h.s.~and r.h.s.~of  (\ref{cocycle3}). For $j=3$ and $j=2$, we have for the l.h.s. of (\ref{cocycle3}):
\begin{equation}\label{lhs1}
  \Omega_{(j) \; \lambda} (x^\rho=a^\rho) \Omega_{(j) \; \rho}(x^\lambda=0) = e^{- i \pi n^{(j)}_{\rho\lambda} H^{(j)}} e^{- i \sum\limits_{\mu\ne \rho}{\pi x^\mu n^{(j)}_{\mu\lambda} \over a^\mu} H^{(j)}} e^{- i \sum\limits_{\mu\ne\lambda}{\pi x^\mu n^{(j)}_{\mu\rho} \over a^\mu} H^{(j)}}, 
\end{equation}
 while the combination of transition functions (\ref{omegabackgrd}) appearing on the r.h.s.  of (\ref{cocycle3}) is
\begin{equation}\label{rhs1}
   \Omega_{(j) \; \rho} (x^\lambda=a^\lambda) \Omega_{(j) \; \lambda}(x^\rho=0) = e^{- i \pi n^{(j)}_{\lambda\rho} H^{(j)}} e^{- i \sum\limits_{\mu\ne \lambda}{\pi x^\mu n^{(j)}_{\mu\rho} \over a^\mu} H^{(j)}} e^{- i \sum\limits_{\mu\ne\rho}{\pi x^\mu n^{(j)}_{\mu\lambda} \over a^\mu} H^{(j)}}.  \end{equation}
Comparing the two expressions (\ref{lhs1}) and (\ref{rhs1}), using the antisymmetry of $n^{(j)}_{\lambda \rho}$, we find that (\ref{omegabackgrd}) obey the cocycle condition
   \begin{equation}\label{cocycle4}
     \Omega_{(j) \; \lambda} (x^\rho=a^\rho) \Omega_{(j) \; \rho}(x^\lambda=0) = e^{i 2 \pi n^{(j)}_{\lambda\rho} H^{(j)}} \Omega_{(j) \; \rho} (x^\lambda=a^\lambda) \Omega_{(j) \; \lambda}(x^\rho=0)~.
\end{equation}
Now,  recalling from (\ref{hmatrices}) that $H^{(3)} = \text{diag}({1\over 3}, {1\over 3}, -1 + {1 \over 3})$ and $H^{(2)} = \text{diag}({1\over 2}, - {1 \over 2})$, we see that (\ref{cocycle4}) produces precisely (\ref{cocycle3}) for $j=2,3$. It is trivial to check that the remaining $U(1)$ condition in  (\ref{cocycle3}) also holds; notice that the $n_{\mu\nu}^{(1)} \in \Z$ factor does not enter the cocycle condition.  

The upshot of the above discussion is that we have shown that (\ref{fractionalinstantons}) are gauge field backgrounds obeying the appropriate cocycle conditions (\ref{cocycle3}) for gauging the $\Z_6^{(1)}$ subgroups, as described in (\ref{gaugingrule}). Having constant field strength and being abelian, they also solve the gauge field equations of motion as well as the scalar $h$ equation of motion with vev set to zero, i.e.~assuming the symmetric phase. The stability of (\ref{fractionalinstantons}) is a consequence of the self-duality, which we impose below.

The topological charges (\ref{topcharge}) of our fractional instantons (\ref{fractionalinstantons}) are now easily computed. In fact, the $Q_{(j)}$ only depend on the transition functions (\ref{omegabackgrd})  but for the constant field strength background we can  compute them 
using the field strength.
 For $j=2,3$, we have 
$F_{(j)} = \sum\limits_{\mu\nu} {\pi dx^\nu \wedge dx^\mu n_{\mu\nu}^{(j)} \over a^\mu a^\nu} H^{(j)}$, 
which yields, using (\ref{hmatrices})
\begin{equation}\label{charge2and3}
Q_{(2)} = {\text{Pf}(n^{(2)}) \over 2},~ Q_{(3)} = {2 \text{Pf}(n^{(3)})  \over 3}~.
\end{equation}
For the $U(1)$, we find instead 
\begin{eqnarray}
F_{(1)} = - \sum\limits_{\mu\nu} {\pi dx^\nu \wedge dx^\mu \over a^\mu a^\nu} \left( {3 n_{\mu\nu}^{(2)} + 2 n_{\mu\nu}^{(3)} \over 6} + n^{(1)}_{\mu\nu}\right)\,,
\label{U1 field strength}
\end{eqnarray}
 yielding
\begin{equation}\label{charge1}
Q_{(1)} = {1 \over 8 \pi^2}\int  F_{(1)} \wedge F_{(1)} = {1 \over 36}\;  \text{Pf}(6 n^{(1)} +3 n^{(2)} + 2 n^{(3)})~.
\end{equation}
As promised earlier, the $U(1)$ topological charge $Q_{(1)}$ is integer for vanishing $n^{(2)}$ and $n^{(3)}$. From (\ref{anomalyb}) we can now immediately compute the $\Delta B$ due to  our fractional instantons
\begin{equation}\label{deltab}
\Delta B = {\text{Pf}(n^{(2)}) \over 2} -  {1 \over 2}\;  \text{Pf}(6 n^{(1)} +3 n^{(2)} + 2 n^{(3)})~.
\end{equation}
It is easy to see that the change of baryon number (\ref{deltab}) is always integer. 

The actions of the fractional instantons (\ref{fractionalinstantons}) can also be computed. The smallest actions of the $SU(2)$ and $SU(3)$ fractional instantons are obtained for solutions on a $\T^4$ obeying (\ref{torussize}).   The   Bogomolny' bound (\ref{bound}) holds after  setting $\text{Pf}(n^{(2)})=\text{Pf}(n^{(3)})=1$ by, for example, taking the only nonzero entries $n_{12}^{(j)}=n_{34}^{(j)}=-n_{21}^{(j)}=-n_{43}^{(j)}=1$ for $j=2,3$:
\begin{eqnarray}
\label{smallest nonabelian actions}
S_{(2)}=\frac{4\pi^2 n_{12}^{(2)}}{g_2^2}\,,\quad S_{(3)}=\frac{16\pi^2 n_{12}^{(3)}}{3g_3^2}\,.
\end{eqnarray}
We note that we have kept $n_{12}^{(j)}$ as a book-keeping device: according to our discussion near eqn.~(\ref{gaugingrule}), we set $n_{12}^{(2)}=n_{12}^{(3)}=1$ upon gauging the full $\Z_6^{(1)}$, we set  $n_{12}^{(2)}=1, n_{12}^{(3)}=0$ when gauging  $\Z_2^{(1)}$, and $n_{12}^{(2)}=0, n_{12}^{(3)}=1$ when gauging $\Z_3^{(1)}$. These  values for $n^{(j)}$, $j=2,3$, give rise to the smallest topological charge for the given gauging. The fractional instantons thus obtained are (anti) self-dual and are necessary stable, as was shown in \cite{vanBaal:1984ar}.

The action of the $U(1)$ instantons, on the contrary, needs to be directly computed starting from the $U(1)$ gauge field background (\ref{fractionalinstantons}). Without loss of generality, we can consider gauge fields in the $1-2$ and $3-4$ planes and take $n_{12}^{(j)} = - n_{21}^{(j)}=n_{34}^{(j)} = - n_{34}^{(j)} = 1$ as the only nonzero entries of $n_{\mu\nu}^{(j)}$, for $j=2,3$, which gives:
\begin{eqnarray}
S_{(1)}=\frac{2\pi^2}{g_1^2} \left[\left(\frac{n_{12}^{(3)}}{3}+\frac{n_{12}^{(2)}}{2}+n_{12}^{(1)}\right)^2+ \left(\frac{n_{12}^{(3)}}{3}+\frac{n_{12}^{(2)}}{2}+n_{34}^{(1)}\right)^2 \right]\,.
\label{smallest abelian action}
\end{eqnarray}
Notice that $U(1)$ solutions with $n_{12}^{(1)}\neq n_{34}^{(1)}$ are not (anti) self-dual. Yet, no unstable modes are expected in this case since the quantum fluctuations about $U(1)$ classical background are always free.

\subsection{Comparing the $\Delta B \ne 0$ rates}

As explained above, different choices of $n_{12}^{(2)}$, $n_{12}^{(3)}$  correspond to gauging different subgroups of $\mathbb Z_6^{(1)}$ (as per (\ref{gaugingrule})) and yield different values of $\Delta B$; the latter also depends on the freedom to add integer $U(1)$ fluxes  $n_{12}^{(1)}$, $n_{34}^{(1)}$. 

Thus, in Table \ref{different choices of n hyper}, we show, for every gauged $1$-form center symmetry, the values of $n_{12}^{(1)}$ and $n_{34}^{(1)}$ that yield the smallest $U(1)$ actions and the corresponding baryon number violation  for  $n_f$ SM generations.\footnote{ 
Notice that the smallest $U(1)$ action in the gauged-$\mathbb Z^{(1)}_6$ case does not result in baryon number violation, and therefore, we needed to consider the next to the smallest action obtained by turning on integer-quantized $U(1)$ fluxes denoted by $n^{(1)}$. The way the scale $\mu_{critical}$ is defined is explained in the text; its value is given for the case of $n_f=3$  generations.}
 
\begin{equation}
\begin{tabular}{|c|c|c|c|c|c|}
\hline
Gauged $1$-form center& $n_{12}^{(1)}$& $n_{34}^{(1)}$ & Smallest $U(1)$ action & $\Delta B$ & $\mu_{\scriptsize\mbox{critical}}$ (GeV) \\\hline\hline
 $\mathbb Z^{(1)}_6$ & $-1$ & $-1$ & $\frac{\pi^2}{9 g_1^2}$ & $0$ & - \\\hline
$\mathbb Z^{(1)}_6$ & $-1 $ & $0 $ & $\frac{13\pi^2}{9 g_1^2}$ & $3 n_f$ & $6\times 10^{34}$   \\\hline
 $\mathbb Z^{(1)}_6$ & $0$ & $-1$ & $\frac{13\pi^2}{9 g_1^2}$ & $3 n_f$ & $6\times 10^{34}$   \\\hline
 $\mathbb Z^{(1)}_3$ & $0$ & $0$ & $\frac{4\pi^2}{9g_1^2}$ & $-2 n_f$ &  $2.7\times 10^{33}$  \\\hline
  $\mathbb Z^{(1)}_2$ & $0$ & $0$ & $\frac{\pi^2}{g_1^2}$ & $-4 n_f$ &  $1.5\times 10^{24}$\\\hline
\end{tabular}\,.
\label{different choices of n hyper}
\end{equation}

Our next task is to compare the vacuum to vacuum amplitude of $\Delta B$ mediated by the fractional-charge instantons with the amplitude of the same quantity due to the weak self-dual BPST instantons. Ideally, one would want to conduct this study at zero temperature on a background manifold $\mathbb M^3\times \R$ that admits both types of instantons. Unfortunately, there is no known background that accommodates analytical solutions of both types. To make matter worse, even on  $\mathbb T^3\times \R$, as we discussed above, analytical solutions of fractional instantons  are still lacking. With these obstacles, our best option is to conduct our study on the simple manifold $\mathbb T^4$. Here, we argue that our treatment gives the correct order of magnitude estimate, which should not be greatly altered by the fine details of the solutions.  First, the action of the self-dual fractional instanton, in the Bogomolny’ bound, depends only on the topological charge and one can proclaim that deforming  $\mathbb T^4$ into $\mathbb T^3\times \R$ (by making one of the cycles very large) cannot change the action.\footnote{In the following, we use the $SU(3)$ solutions with topological charge $\frac{2}{3}$ instead of the solutions with the minimum charge $\frac{1}{3}$ described by 't Hooft  \cite{tHooft:1981nnx}, see Footnote \ref{footnoteabelian}. We also assume that the $U(1)$ topological charge and action of the constant-field solution (\ref{U1 field strength}) does not change dramatically during this process, i.e., when we make one of the $\T^4$ cycles very large.} Second, although it is true that there is no solution of self-dual BPST instanton on $\mathbb T^4$ with zero twists \cite{Braam:1988qk}, adding a twist removes the obstruction to the existence of the solution.\footnote{See \cite{vanBaal:2000zc} for a discussion. Simply, if one starts with a self-dual BPST instanton on $\mathbb R^4$, then deforming it to $\mathbb T^4$ will be obstructed since the latter manifold enjoys a continuous degeneracy in the moduli space. This degeneracy, however, will be lifted when we apply the twist.  Yet, from a practical point of view, one can get very close to self-duality even in the absence of twists.} In our case the twists come about naturally, since we need them anyway to obtain the fractional solutions. This is a win-win situation for both the BPST and fractional solutions. Again, one can imagine a process where we take one of the cycles of $\T^4$ large enough to obtain BPST solutions on $\mathbb T^3\times \R$.  A third important point is that  baryon number violation is governed mainly by anomalies. The latter depend only on the cohomology classes and not on the specific background manifold, and thus, we expect the computations on $\mathbb T^4$ or $\mathbb T^3\times \mathbb R$  yield the same estimates. 

Given all the above caveats and their suggested resolutions, we proceed to our computations. We begin with $\mathbb T^4$ of a size much smaller than the inverse TeV scale and, therefore, we consistently set the Higgs vev to zero and use the solutions (\ref{fractionalinstantons}) and the corresponding action for the fractional instantons. As we discussed above, finding BPST instanton (approximate) solutions on $\mathbb T^4$ should also be possible, with a scale modulus cutoff by the torus size.  We then declare that the actions of the BPST and fractional instantons will not be greatly modified as we take one of the $\mathbb T^4$ cycles large enough, while keeping the $\mathbb T^3$ size below the inverse TeV scale, such that we approach the zero-temperature limit $\mathbb T^3\times \R$. In this limit, the small size of  $\mathbb T^3$ will set the energy scale of the problem.  We also give the SM fermions periodic boundary conditions along the spatial directions $\mu=1,2,3$.  Effectively, the problem is reduced to quantum mechanics, where there are two types of nonperturbative solutions that compete with each other to win  the baryon-number-violation race. The action of the solution(s) is the highest of the barriers between the different vacuum components, and the solution with the smallest action will more effectively violate the baryon number. In this regime the  schematic form of 't Hooft vertex in the background of fractional instantons is given by (we consider the contribution from one family of quarks and leptons and the field insertions $q_L$ etc. are the fermion zero modes)
 \begin{eqnarray}
 {\cal T}_{\scriptsize \mbox{fraction}}\sim e^{-\left(S_{(1)}+S_{(2)}+S_{(3)}\right)}(q_L)^{{\cal I}_{q_L}} (l_L)^{{\cal I}_{l_L}} (\tilde e_R)^{{\cal I}_{\tilde e_R}} (\tilde u_R)^{{\cal I}_{\tilde u_R}}  (\tilde d_R)^{{\cal I}_{\tilde d_R}}\,, 
 \label{fractional thooft vertex}
 \end{eqnarray}
 where ${\cal I}_{q_L}$, etc. are the Dirac indices:
 \begin{eqnarray}
 \nonumber
 {\cal I}_{q_L}&=&2Q_{(3)}+3Q_{(2)}+6Q_{(1)}\,,\quad  {\cal I}_{l_L}=Q_{(2)}+18 Q_{(1)}\,,\quad {\cal I}_{\tilde e_R}=36Q_{(1)}\,,\\
 {\cal I}_{\tilde u_R}&=&Q_{(3)}+48 Q_{(1)}\,,\quad  {\cal I}_{\tilde d_R}=Q_{(3)}+12Q_{(1)}. \label{indices}
 \end{eqnarray}
 The indices are obtained from the index theorem, which gives the number of the Weyl zero modes in the background of the fractional-charge instanton. The indices (\ref{indices}), like $\Delta B$ of (\ref{deltab}), are always integer, for fermions whose transition functions are consistent with the cocycle conditions (\ref{cocycle1}, \ref{cocycle3}) (this can also be explicitly checked). 
 Inserting the vertex into the SM partition function and applying a $U(1)_B$ global transformation gives $\Delta B= {1 \over 3}({\cal I}_{q_L}- {\cal I}_{\tilde u_R}- {\cal I}_{\tilde d_R})=Q_{(2)}-18 Q_{(1)}$, which  is exactly the result from the anomaly equation (\ref{anomalyb}). Let us compare (\ref{fractional thooft vertex})  to 't Hooft vertex from the weak BPST instantons that give the same $\Delta B$ (per family)
 \begin{eqnarray}
 {\cal T}_{\scriptsize\mbox{BPST}}\sim e^{-|\Delta B| S_{\scriptsize\mbox{BPST}}} \left(q_L^3 l_L\right)^{|\Delta B|}\,.
 \end{eqnarray}

Bearing in mind that the actions are  coupling constant dependent,  the sum of the actions  $S_{(1)}+S_{(2)}+S_{(3)}$ could become smaller than the action of the  BPST instantons (times $\Delta B$), which, in turn, results in a larger amplitude for the fractional-instanton-mediated process. Using (\ref{smallest nonabelian actions}), (\ref{smallest abelian action}), and $S_{\scriptsize\mbox{BPST}}=\frac{8\pi^2}{g_2^2}$,   the condition $S_{(1)}+S_{(2)}+S_{(3)}< |\Delta B| S_{\scriptsize\mbox{BPST}}$ can be rewritten as:
\begin{eqnarray}
\nonumber
\frac{16\pi^2 n_{12}^{(3)}}{3g_3^2(\mu)}+\frac{2\pi^2}{g_1^2(\mu)} \left[\left(\frac{n_{12}^{(3)}}{3}+\frac{n_{12}^{(2)}}{2}+n_{12}^{(1)}\right)^2+ \left(\frac{n_{12}^{(3)}}{3}+\frac{n_{12}^{(2)}}{2}+n_{34}^{(1)}\right)^2 \right]<\frac{8\pi^2}{g_{2}^2(\mu)}\left(|\Delta B|-\frac{n_{12}^{(2)}}{2}\right)\,.\\
\label{main inequality}
\end{eqnarray}
As above, $n_{12}^{(2)}$ and $n_{12}^{(3)}$ (taken to be either $0$ or $1$) are used as a book-keeping device to track the gauging of $\mathbb Z_{6}^{(1)}$ or subgroups of it, while $n_{12}^{(1)}$ and $n_{34}^{(1)}$ are as in Table \ref{different choices of n hyper}. 

As we emphasized in the inequality (\ref{main inequality}), the coupling constants $g_{1,2,3}$ depend on some energy scale $\mu$. Above the electroweak scale, the Higgs is in the symmetric phase and the only scale in the problem is the $\mathbb T^3$ size, which works as an IR cutoff and  sets the energy scale $\mu$.  To determine the size of  $\T^3$  that satisfies the inequality (\ref{main inequality}), we run the coupling constants from the weak scale (which is taken to be the $Z$-boson mass $M_Z$) to the scale $\mu$. The running is governed by:
\begin{eqnarray}
\nonumber
\frac{8\pi^2}{g_i^2(\mu)}&=&\frac{8\pi^2}{g_i^2(M_Z)}+b_i\log \left(\frac{\mu}{M_Z}\right)\,,\quad i=1,2,3\\
b_1&=&-(80 n_f+6 n_H)\,,\quad b_2=\frac{22}{3}-\frac{4}{3}n_f-\frac{n_H}{6}\,,\quad b_3=11-\frac{4}{3}n_f\,,
\label{running of couplings}
\end{eqnarray}
 for $n_f$ families of particles  and $n_H$ Higgs doublets. Notice that while  $b_2$ and $b_3$ are standard (see, e.g., \cite{Martin:1997ns}),  the $U(1)$ coupling constant  $g_1$ as well as its $\beta$-function are related to the ``standard" hypercharge via $g_1=g_Y/6$. We also use $g_3^2 (M_Z)/4\pi\cong 0.118$, $g_{EM}^2 (M_Z)/4\pi \cong 1/128$, $g_Y (M_Z)=g_{EM}(M_Z)/\cos\theta_W(M_Z)$, $g_2 (M_Z)=g_{EM}(M_Z)/\sin\theta_W(M_Z)$, and $\sin^2\theta_W(M_Z)\cong 0.23$, where $\theta_W(M_Z)$ is the Weinberg angle. 
 
Using the SM particle content, setting $n_f=3$ and $n_H=1$, and making use of (\ref{running of couplings}) in (\ref{main inequality}), we solve for the critical energy scale $\mu_{\scriptsize\mbox{critical}}$ at which the effects of the fractional-charge instantons surpass the BPST ones. We list the values of  $\mu_{\scriptsize\mbox{critical}}$ in the last column in Table \ref{different choices of n hyper}. Clearly, all  values of  $\mu_{\scriptsize\mbox{critical}}$  are super-Planckian for the gauging of $\mathbb Z_{6}^{(1)}$ or its subgroups.  This is attributed to the fact that the hyper-charge coupling constant is extremely small at the weak scale. One can push $\mu_{\scriptsize\mbox{critical}}$   into sub-Planckian values provided that extra matter charged under $U(1)$ participates in the running.\footnote{It is important to stress that any extra charged matter must respect the modified cocycle conditions (\ref{cocycle3}). We also assume that such new particles have mass above the TeV scale. As an example, we can assume  $\sim 60$  extra scalar particles of mass $\sim 10^5$ GeV,  which are sterile under $SU(2)$ and $SU(3)$ but have a charge $6$ under $U(1)$. They push the critical scale down to $\mu_{\scriptsize \mbox{critical}}\sim 10^{13}$ GeV for $G_{SM}^{2}$. Even at this energy, the $U(1)$ coupling is still in the perturbative regime.}

Finally, we might be interested in the energy scale at which the  fractional instantons become more important than  BPST solutions with minimum topological charge $Q_{(2)}=1$. In this case we simply set $|\Delta B|=1$ in the inequality (\ref{main inequality}) to find that  $\mu_{\scriptsize\mbox{critical}}$  are not dramatically bigger than the values in Table \ref{different choices of n hyper}. 
 
\section{Cosmological bounds and discussion}
 
 The strongest cosmological constraints on the size of a $\T^3$  with sides of order $L$ come from the CMB and indicate that $L > {\cal{O}}(few) \times L_0$, where $L_0 \sim 14$ Gpc is the radius of the last scattering surface, see e.g. \cite{Levin:1998qq,Bond:1997ym,Aslanyan:2013lsa}. If compactness of the space is a necessary condition to make sense of the fractional-charge instantons, which is the conservative perspective that we have been following throughout our analysis (however, see the discussion below), then these instantons could have played a role in the early Universe provided that there is extra charged matter and the size of $\T^3$ was smaller than or equal to the horizon size when the Universe was hot and dense. 
 
To this end, we trace back the size of $\T^3$ before the time of last scattering,  when the Universe was deep in the radiation-dominated era at temperature $T$ and the instanton effects are expected to be important. We use the Friedmann equation of a single-component Universe $3M_P^2H^2(t)=\frac{\pi^2}{15}n_*T^4$ ($M_P$ is the reduced Planck mass, $H(t)$ is the Hubble parameter, and $n_*$ is the number of effective degrees of freedom), along with the fact that the scale factor behaves as $a\propto 1/\sqrt{H(t)}$. Recalling that the horizon size is $L_H(t)=H^{-1}(t)$,  we can use this information to obtain the horizon size, the  physical size $L_{\T^3}(T)$ of $\T^3$, and the  ratio $ \frac{L_{\T^3}(T)}{L_H(T)}$ at temperature $T$:
 \begin{eqnarray}
L_H(T)\sim\frac{M_P}{T^2}\,,\quad L_{\T^3}(T)=  {\cal{O}}(few)\frac{M_P}{T_l T}\,,\quad  \frac{L_{\T^3}(T)}{L_H(T)}= {\cal{O}}(few) \times\frac{T}{T_l}\,,
 \end{eqnarray}
 where $T_l$ is the CMB temperature at the time of last scattering. Since we are interested in temperature much higher than $T_l$, we immediately see that the $\T^3$ was always bigger than the horizon size. One might immediately run into the conclusion that such fractional solutions should be dismissed since their size has always been bigger than the horizon. This, however, is a very conservative view that needs to be carefully examined in a future study. Moreover, it is conceivable that such solutions can be found at a scale cut off by the horizon. 
  
Assuming one can make sense of the fractional instantons on scales smaller than or equal to the horizon size, then things become more interesting at temperature above the electroweak scale. Now, the electroweak symmetry is restored and there is no barrier between the pre-vacua of $SU(2)$. In this case the baryon and lepton number violation happens in units of $3$, one per family, at a rate set by temperature $\sim T^4$.  If there is no extra matter charged under $U(1)$, then our conclusion from the previous section implies that fractional instantons did not play any role during the thermal history of the Universe. If, however, there is extra matter that makes the fractional instantons more important at some critical energy scale $\mu_{\scriptsize\mbox{critical}}<M_P$, then at temperature $T \gtrsim \mu_{\scriptsize\mbox{critical}}$ the barrier due to the fractional instantons is removed and the baryon and lepton number violation is expected to proceed at similar rates $\sim T^4$  in units of $\Delta B>3$, as given in Table \ref{different choices of n hyper}.

{\bf {\flushleft{Acknowledgments:}}} M.A. is supported by STFC through grant ST/T000708/1.   EP is supported by a Discovery Grant from NSERC.

\bigskip

  \bibliography{RefSMglobal1.bib}
  
  \bibliographystyle{JHEP}
  \end{document}